\documentclass[12pt]{article}
\usepackage{amsmath,amssymb}
\usepackage{hyperref}
\numberwithin{equation}{section}

\newcommand{\Ncal}{\mathcal{N}}

\newcommand{\del}{\partial}

\newcommand{\Fcal}{\mathcal{F}}
\newcommand{\Ocal}{\mathcal{O}}

\begin{document}

%%%%%%%%%%%%%%%%%%%%%%%%%%%%%%%%%%%%%%%%%%%%
\thispagestyle{empty}
\begin{flushright}
OU-HET 749
\end{flushright}
\vskip3cm
\begin{center}
{\LARGE {\bf Expectation values of chiral primary operators in holographic interface CFT}}
\vskip1.5cm
{\large 
Koichi Nagasaki\footnote{nagasaki [at] het.phys.sci.osaka-u.ac.jp}
and\hspace{2mm} Satoshi Yamaguchi\footnote{yamaguch [at] het.phys.sci.osaka-u.ac.jp}
}
\vskip.5cm
{\it Department of Physics, Graduate School of Science, 
\\
Osaka University, Toyonaka, Osaka 560-0043, Japan}
\end{center}

%%%%%%%%%%%%%%%%%%%%%%%%%%%%%%%%%%%%%%%%%%%%
\vskip2cm
\begin{abstract}
We consider the expectation values of chiral primary operators in the presence of the interface in the 4 dimensional $\Ncal=4$ super Yang-Mills theory.  This interface is derived from D3-D5 system in type IIB string theory.  These expectation values are computed classically in the gauge theory side.  On the other hand, this interface is a holographic dual to type IIB string theory on AdS$_5\times$S$^5$ spacetime with a probe D5-brane.  The expectation values are computed by GKPW prescription in the gravity side.  We find non-trivial agreement of these two results: the gauge theory side and the gravity side.
\end{abstract}

%%%%%%%%%%%%%%%%%%%%%%%%%%%%%%%%%%%%%%%%%%

\newpage
\tableofcontents
%%%%%%%%%%%%%%%%%%%%%%%%%%%%%%%%%%%%%%%%%%%%
\section{Introduction and summary}
The AdS/CFT correspondence \cite{Maldacena:1997re} is an interesting duality between a gravity theory and a gauge theory.  However it is very difficult to check this duality since unprotected quantities are calculable only in the small 't Hooft coupling $\lambda$ regime in the gauge theory side, while they are 
calculable in the large $\lambda$ regime in the gravity side.

There are several ways to overcome this difficulty.  One of them is to introduce another large parameter as in \cite{Berenstein:2002jq}.   In \cite{Berenstein:2002jq}, the R-charge $J$ (the angular momentum in the gravity side) has been taken to be large and the effective expansion parameter has become $\lambda/J^2$.  By virtue of this change of the effective coupling, the conformal dimension of such operators have been successfully compared to the energy of the stringy excited states in the pp-wave geometry.  This result has given a non-trivial evidence of the AdS/CFT correspondence.  Other examples of similar phenomena are found in surface operators \cite{Drukker:2008wr} (see also \cite{Koh:2008kt}) and the interface \cite{Nagasaki:2011ue}.

An interface is a wall in the space-time which connect two different (or the same) quantum field theories. A partial list of related references are \cite{Sethi:1997zza,Ganor:1997jx,Kapustin:1998pb,Karch:2000gx,DeWolfe:2001pq,Bachas:2001vj,Bak:2003jk,D'Hoker:2006uu,D'Hoker:2006uv,Gomis:2006cu,D'Hoker:2007xy,D'Hoker:2007xz,Gaiotto:2008sa,Gaiotto:2008sd,Gaiotto:2008ak}. See also \cite{Kirsch:2004km} and references there in.  The interface considered in this paper is so called a ``Nahm pole'' which connects SU$(N)$ gauge theory and SU$(N-k)$ gauge theory.  The boundary condition is determined by the fuzzy funnel solution \cite{Constable:1999ac}.  In this interface a parameter $k$ is introduced, and taken to be large in this paper as done in \cite{Nagasaki:2011ue}.  This interface is described by the intersecting D3-D5 system where $k$ D3-branes end on the D5-brane.  Thus the gravity dual is given by the near horizon limit of the supergravity solution for the D3-branes with the probe D5-brane with $k$ units of magnetic flux \cite{Karch:2000gx}.

In this paper we study the expectation values of chiral primary operators in the presence of the above interface.  In the gauge theory side the expectation values are evaluated by just substituting the classical solution of the fuzzy funnel solution.  On the other hand they are calculated by GKPW prescription \cite{Gubser:1998bc,Witten:1998qj}.  Usually these two results cannot be compared to each other because the gauge theory result is only valid in the small $\lambda$ regime, while the gravity result is only valid in the large $\lambda$ regime.  However in our case we can take $k\to\infty$ limit and make $\lambda/k^2$ small even if $\lambda$ is large in the gravity side.  In this limit we find perfect agreement between the gauge theory result and the gravity result.  This is a quite non-trivial evidence of the AdS/CFT correspondence.

The construction of this paper is as follows. In section \ref{sec:gauge}, we review the 4-dimensional $\Ncal=4$ SYM theory and the interface, and show the calculation of the expectation values of the chiral primary operators in the presence of the interface.  In section \ref{sec:gravity}, we turn to the calculation in the gravity side using GKPW prescription. In section \ref{sec:comparison}, the above two results are compared and the perfect agreement is found in the leading order.  The next-to-leading term is predicted from the gravity side.

%%%%%%%%%%%%%%%%%%%%%%%%%%%%%%%%%%%%%%%%%%%%
\section{Gauge theory side}\label{sec:gauge}
We consider the 4-dimensional $\mathcal{N}=4$ supersymmetric Yang-Mills theory in this section.
We review the action of this theory and classical solutions. After that we calculate the expectation values of the chiral primary operators in the presence of the interface.

\subsection{Fields and Action}
We consider here the $\mathcal{N}$=4 super Yang-Mills theory with the gauge group $SU(N)$. We use the same convention as \cite{Nagasaki:2011ue}. The action is given by
\begin{equation}
S=\frac{2}{g^2}\int d^4x \text{tr}\left[
-\frac{1}{4}F_{\mu\nu}F^{\mu\nu}
-\frac{1}{2}D_\mu\phi_iD^\mu\phi^i
+\frac{i}{2}\bar{\psi}\Gamma^\mu D_\mu \psi
+\frac{1}{2}\bar{\psi}\Gamma^i [\phi_i,\psi]
+\frac{1}{4}[\phi_i,\phi_j][\phi^i,\phi^j]
\right],
\end{equation}
where $F_{\mu\nu}$,\ $\mu=0,\cdots,3$, are the field strength of the gauge field $A_{\mu}$, which is expressed as $F_{\mu\nu}=\del_{\mu}A_{\nu}-\del_{\nu}A_{\mu}-i[A_{\mu},A_{\nu}]$.  While $\psi$ is a fermion field and $\phi_i$, $i=4,\cdots,9$, are scalar fields.  All these fields are in the adjoint representation of SU$(N)$, in other words, $N\times N$ hermitian traceless matrices.
These scalar fields play a crucial role in the one-point function we want to calculate in this paper.

This action is invariant under the following supersymmetry transformation with the spinor parameter $\epsilon$.
\begin{eqnarray}
&&\delta A_\mu = i\bar\epsilon\Gamma_\mu \psi,\\
&&\delta \phi_i = i\bar\epsilon\Gamma_i \psi,\\
&&\delta\psi = \frac{1}{2}F_{\mu\nu}\Gamma^{\mu\nu}\epsilon+D_\mu\phi_i\Gamma^{\mu i}\epsilon -\frac{i}{2}[\phi_i,\phi_j]\Gamma^{ij}\epsilon.
\end{eqnarray}

\subsection{Interface}\label{FuzzyFunnelSol}
We introduce here a wall-like object called interface. 
%We also use in this paper the same convention as \cite{Nagasaki:2011ue}.
This object separates a whole space into two regions where gauge theories with different gauge groups live. One has gauge group SU$(N)$ and the other has SU$(N-k)$.
This interface is defined by a classical solution known as a fuzzy funnel solution \cite{Constable:1999ac}.
This solution plays a crucial role in our calculation.
The interface is defined by a boundary condition between two different gauge theories and leads to a non-trivial classical vacuum solution.
\begin{equation}\label{ClVacuumsol}
A_\mu=0,\hspace{1cm}
\phi_i=\phi_i(x_3),\:(i=4,5,6),\hspace{1cm}
\phi_i=0,\:(i=7,8,9).
\end{equation}
The solution $\phi_i=\phi_i(x_3),\:(i=4,5,6),$ is called a fuzzy funnel solution \cite{Constable:1999ac}. 
%\subsection{Fuzzy-Funnel solution}
The solution of scalar fields are given by
\begin{equation}
\phi_i=-\frac{1}{x_3}t_i\oplus 0_{(N-k)\times(N-k)}\hspace{1cm} (x_3> 0),
\end{equation}
where $t_i,\: i=4,5,6$, are $k\times k$ matrices which denote the generators of SU(2) algebra of the $k$-dimensional irreducible representation.
%$k$ is a parameter which corresponds to the number of the D3 branes which end on the D5 brane.
The following relation is useful for our calculation.
\begin{equation}\label{ConvenientRelations}
\phi_4^2+\phi_5^2+\phi_6^2=\frac{1}{4x_3^2}(k^2-1){1}_{k\times k}\oplus{0}_{(N-k)\times(N-k)}.
\end{equation}

%\subsection{Chiral primary}
\subsection{One-point function}
In this section we consider the one-point functions of chiral primary operators.  The chiral primary operators are defined as
\begin{equation}
\mathcal{O}_\Delta(x):=\frac{(8\pi^2)^{\Delta/2}}{\lambda^{\Delta/2}\sqrt{\Delta}}C^{I_1I_2\cdots I_\Delta}\text{Tr} (\phi_{I_1}(x)\phi_{I_2}(x)\cdots \phi_{I_\Delta}(x)),
\end{equation}
where $\Delta$ denotes the conformal dimension and $C^{I_1I_2\cdots I_\Delta}$ is a traceless symmetric tensor normalized as $C^{I_1I_2\cdots I_\Delta}C^{I_1I_2\cdots I_\Delta}=1$.
The normalization of the operator is determined so that the two point function without interface becomes
\begin{align}
 &\langle \mathcal{O}_\Delta(x)\mathcal{O}_\Delta(y)\rangle=\frac{1}{|x-y|^{2\Delta}}.
\end{align}
 See \cite{Lee:1998bxa} for the detail.

We would like to calculate the one-point function of this operator.  Let us insert this operator at a point $x_3=\xi$ and consider the expectation value $\langle \mathcal{O}_{\Delta}(\xi)\rangle$.
For calculating the classical expectation value of this operator we substitute the fuzzy funnel solution introduced in the above section \ref{FuzzyFunnelSol}. 
Since our fuzzy funnel solution preserves SO$(3)\times$SO$(3)$ symmetry, only SO$(3)\times$SO$(3)$ invariant chiral primary operators can have non-vanishing expectation values. As shown in Appendix \ref{app:harmonics}, $\Delta$ must be even and is denoted as $\Delta=2\ell$.  Moreover there is only one such chiral primary operator for each $\Delta=2\ell,\ \ell=0, 1,2,3,\cdots$. 

The traceless symmetric tensors $C^{I_1\cdots I_{\Delta}}$ are related to the spherical harmonics (see Appendix \ref{app:harmonics}).
\begin{equation}
C^{I_1I_2\cdots I_\Delta}x_{I_1}\cdots x_{I_\Delta}=Y_\ell (\psi),\hspace{1cm}
\sum_{i=4}^{6} x_i^2=\sin^2\psi,\hspace{1cm}
\sum_{j=7}^{9} x_j^2=\cos^2\psi
\end{equation}
Spherical harmonics is expressed as eq.~\eqref{Cl}
\begin{align}
Y_\ell (\psi) 
= C_\ell F(-\ell,\ell+2,\frac{3}{2};\cos^2\psi)=C_\ell (1+\cos^2\psi P(\cos^2\psi)),\label{inhomogeneous}
\end{align}
where $P(\cos^2\psi)$ is an inhomogeneous polynomial of $\cos^2\psi$.
The normalization $C_{\ell}$ is determined so that $C^{I_1I_2\cdots I_\Delta}C^{I_1I_2\cdots I_\Delta}=1$ is satisfied, or equivalently eq.~\eqref{Ynormalization}.
We can express this spherical harmonics by a homogeneous polynomial of $\sin^2 \psi$ and $\cos^2 \psi$. This is because if we have a inhomogeneous term, we can replace $1$ by some power of $\sin^2\psi+\cos^2\psi$.  In particular we can replace the first term $1$ in the paren in eq.~\eqref{inhomogeneous} by $(\sin^2\psi+\cos^2\psi)^{\ell}$ and get homogeneous expression
\begin{align}
Y_{\ell}= C_\ell (\sin^{2\ell}\psi+\cos^2\psi\; Q(\sin^2\psi,\cos^{2}\psi)),
\end{align}
where $Q(\sin^2\psi,\cos^{2}\psi)$ is a homogeneous polynomial of $\sin^2\psi$ and $\cos^2\psi$.
Then replacing $\sin^2\psi$ by $\sum_{i=4}^{6}\phi_i^2$ and $\cos^2\psi$ by $\sum_{j=7}^{9}\phi_j^2$, we obtain the relation\footnote{Precisely speaking the right hand side is symmetrized product.}
\begin{equation}
C^{I_1\cdots I_\Delta}\phi_{I_1}\cdots \phi_{I_\Delta}=C_\ell \left\{\left(\sum_{i=4}^{6}\phi_i^2\right)^\ell +\left(\sum_{j=7}^{9}\phi_j^2\right) Q\left(\sum_{i=4}^{6}\phi_i^2,\sum_{j=7}^{9}\phi_j^2\right)\right\}.
\end{equation}
Substituted the solution \eqref{ClVacuumsol}, all terms except the first one vanish since $\phi_7=\phi_8=\phi_9=0$.
Using the relations \eqref{ConvenientRelations} we obtain the following result.
\begin{eqnarray}
\left<\mathcal{O}_{2\ell}(\xi)\right>_\text{classical}
&=& \frac{(8\pi^2)^{\Delta/2}}{\lambda^{\Delta/2}\sqrt{\Delta}}C_\ell \text{Tr}\left[\left(\frac{1}{4\xi^2}(k^2-1)\right)^\ell \bold{1}_{k\times k}\right]\nonumber\\
&=& C_\ell\frac{(2\pi^2)^\ell}{\sqrt{2\ell}\lambda^\ell}(k^2-1)^{\ell}k\frac{1}{\xi^{2\ell}}.
\label{gauge-result}
\end{eqnarray}
The behavior $1/\xi^{2\ell}$ is determined by the conformal symmetry and does not change by the quantum correction. The non-trivial part is the coefficient, which will change by the quantum correction.  We compare this result with the gravity side calculation.  
%%%%%%%%%%%%%%%%%%%%%%%%%%%%%%%%%%%%%%%%%%%%
\section{Gravity side}\label{sec:gravity}
%In this section we show the result from gravity side.
%In the previous section we calculated the expectation values of chiral primary operators in the classical limit in the gauge theory side.
%We will calculate the corresponding quantity in the gravity side in this section and compare these two results in the next section.
In this section we calculate the expectation values of the chiral primary operators in the gravity side.
The AdS/CFT correspondence is a duality between the $\mathcal{N}=4$ super Yang-Mills theory we discussed in the previous section and type IIB superstring theory on AdS$_5\times$S$^5$.  How this gravity side is modified when the interface is inserted?
%These theories are two different views of a D3 brane configuration. \\
%How do we construct the interface we discussed in gauge side?
The object which corresponds to our interface is a probe D5-brane with $k$ units of  magnetic flux \cite{Karch:2000gx}.  This gravity dual is obtained by the following way.
We consider a D5-brane where $k$ D3-branes end.
%We refer the number of D3 branes ending on the D5 brane as $k$.
Then SU($N$) gauge theory is realized in the side where there are $N$ D3 branes and SU($N-k$) gauge theory is realized in the other side as low energy effective theories.  This D5-brane is pulled by $k$ D3-branes which end on it and become funnel shape with $k$ units of magnetic flux.  If we consider the supergravity solution of D3-branes and take the near horizon limit, we obtain the gravity dual mentioned above.
%Hence this D5 brane should be regarded as corresponding object of interface in gravity side.\\

Here we make a remark on the value $k$. Although we take $k$ large, it is still much smaller than $N$ in order not to modify the supergravity background.
%%%%%%%%%%%%%%%%%%%%%%%%%%%%%
\subsection{The Gubser-Klebanov-Polyakov-Witten relation}
%AdS/CFT correspondence suggest the following relation.
%\begin{equation}
%Z_\text{CFT}=Z_\text{string}
%\end{equation}
%where the left hand side is the partition function of the conformal field theory and the right hand side is that of the superstring.

The correlation functions in the AdS/CFT correspondence are calculated by GKPW prescription \cite{Gubser:1998bc,Witten:1998qj}.  Due to GKPW there is one-to-one correspondence between local operators in the gauge theory and fields in the gravity theory.  Let $\Ocal$ be a scalar operator in the gauge theory, and $s$ be the scalar field in the gravity theory which corresponds to $\Ocal$.  GKPW claims that the relation
\begin{equation}
\left<\text{e}^{\int d^4xs_0(x)\mathcal{O}(x)}\right>_\text{CFT}
=
\text{e}^{-S_\text{cl}(s_0)}
\end{equation}
is satisfied in the classical gravity limit.  In this equation $s_0$ is a boundary condition of $s$ up to a certain factor, $S_\text{cl}(s_0)$ is the action evaluated by the classical solution with the boundary condition given by $s_0$.
%
%Now let $s_0(x)$ be the value of scalar field on the boundary of AdS$_5$. This field is coupled with primary operators $\mathcal{O}$: $\int s_0\mathcal{O}$.
%In order to calculate the correlation functions $\left<\mathcal{O}(x_1)\cdots\mathcal{O}(x_n)\right>$ we define the generating functional $Z_\text{CFT}=\left<\int \exp (s_0\mathcal{O})\right>_\text{CFT}$. \\
%In gravity theory let $S_\text{cl}$ be the classical supergravity action. The supergravity partition function can be calculated as $Z_\text{S}\approx \text{e}^{-S_\text{cl}(s_0)}$.
%Then conformal field theory and supergravity are related as
%This relation is known as the Gubser-Klebanov-Polyakov-Witten relation, called GKPW for short. \cite{Gubser:1998bc} \cite{Witten:1998qj}
%
Using this relation the one-point function is calculated as follows.
\begin{equation}
\left<\mathcal{O}(x)\right>
%=\frac{\delta}{\delta s_0(x)}\exp\{-S\}\Big|_{s_0=0}
	=\left. \frac{-\delta S_\text{cl}(s_0)}{\delta s_0(x)}\right|_{s_0=0}.
\label{GKPW1pt}
\end{equation}
We employ the normalization $\langle 1 \rangle=1$.  

If no interface or other defects are inserted, this one-point function vanishes due to the conformal invariance.   In terms of the gravity theory, this one-point function vanishes since the background is a solution of the equation of motion and thus any variation of the action vanishes at this background.  In our case this one-point function does not vanish in general because the interface is inserted as we have seen in the previous section.  In the gravity side, this one-point function does not vanish because we have, in addition to the supergravity, a probe D5-brane which gives non-vanishing contribution.

%%%%%%%%%%%%%%%%%%%%%%%%%%%%%%%
\subsection{Background}
We consider here type IIB superstring theory as the gravity theory.
The near horizon limit of the supergravity solution of $N$ coincident D3-branes is AdS$_5\times$S$^5$. 
The coordinates of AdS${}_5$ are denoted by $y,x^\mu,\mu=0,1,2,3$.
% and $x^\mu,\mu=4,\cdots,9$ on S$^5$.
%Due to the existence of the D5-brane our spacetime is curved.
The metric on this space is given by
\begin{equation}
ds^2_\text{AdS$_5\times$S$^5$}=\frac{1}{y^2}(dy^2+\eta_{\mu\nu}dx^\mu dx^\nu)+ds^2_\text{S$^5$}.
\end{equation}
In this paper we choose the unit in which the radius of AdS${}_5$ is $1$.  Thus 
the string coupling constant $g_s$ and the slope parameter $\alpha'$ are related as
\begin{equation}
\lambda:=4\pi g_s N =\alpha'^{-2}. \label{unit}
\end{equation}
Furthermore the RR 4-form is also excited
\begin{equation}
C_4=-\frac{1}{y^4}dx^0dx^1dx^2dx^3+\cdots.
\end{equation}

%\subsection{D5 brane solution}
In addition to the D3-brane configuration discussed earlier, we introduce a D5-brane in order to study the corresponding theory of the interface CFT.
The D5-brane action is the usual DBI+WZ action.
\begin{equation}
S=T_5 \int d^6\zeta\sqrt{\det (G+\Fcal)}+iT_5\int \mathcal{F}\wedge C_4,
\end{equation}
where $T_5=(2\pi)^{-5}\alpha'^{-2}g_s^{-1}$ is the tension of the D5-brane, $\zeta$'s are the world-volume coordinates, $G$ and $\Fcal$ denote the induced metric and the field strength of the world-volume gauge field respectively. 
%\begin{equation}
%T_5=\frac{1}{(2\pi)^5\alpha'^3g_s}, \hspace{1cm}
%\alpha'=\frac{1}{\lambda}, \hspace{1cm}
%4\pi g_sN\alpha'^2=1
%\end{equation}

The AdS${}_4\times$S${}^2$ solution is obtained by \cite{Karch:2000gx}.   We use the convention of \cite{Nagasaki:2011ue}.  AdS$_{4}$ part is embedded in AdS${}_5$ and expressed by the equation
\begin{equation}\label{D5inAdS}
x_3=\kappa y
\end{equation}
with a constant parameter $\kappa$.  S$^2$ is embedded in S${}^5$ as a great sphere.
We denote world-volume coordinates of D5 by $(y,x_0,x_1,x_2,\theta,\phi)$; $(y,x_0,x_1,x_2)$ are coordinates of AdS${}_4$ and $(\theta,\phi)$ are ones of S${}^2$.  The induced metric and the gauge field are summarized by a matrix
$H=G+ \mathcal{F}$. $H$ takes the following form in this solution.
\begin{equation}
H=
\left(\begin{array}{cccc|cc}
(1+\kappa^2)y^{-2} &  &  &  &  &  \\
  & y^{-2} &  &  &  &  \\
  &  & y^{-2} &  &  &  \\
  &  &  & y^{-2} &  &  \\\hline
  &  &  &  & 1 & -\kappa\sin\theta \\
  &  &  &  & \kappa\sin\theta &\sin^2\theta \end{array}\right).
\end{equation}
Actually the parameter $\kappa$ is related with $k$ as $\kappa = \frac{\pi}{\sqrt{\lambda}}k$.
%Since the equation of motion of D5-brane  
%The D5-brane is expanded as \eqref{D5inAdS} and wrapped in S$^2$ of S$^5$.

\subsection{One-point function from gravity theory}
Now let us turn to the calculation of the one point function.  The scalar fields which correspond to the chiral primary operators are identified in \cite{Kim:1985ez,Lee:1998bxa}.  These scalar fields come from the fluctuation of the metric and the RR 4-form as
\begin{align}
&h^\text{AdS}_{\mu\nu} 	=-\frac{2\Delta(\Delta-1)}{\Delta+1}sg_{\mu\nu}+\frac{4}{\Delta+1}\nabla_\mu\nabla_\nu s ,\label{Fluc1}\\
&h^\text{S}_{\alpha\beta}	=2\Delta sg_{\alpha\beta} , \label{Fluc2}\\
&a^\text{AdS}_{\mu\nu\rho\sigma}=4i\sqrt{g^\text{AdS}}\epsilon_{\mu\nu\rho\sigma\eta}\nabla^\eta s,\label{Fluc3}
\end{align}
where $h^\text{AdS}_{\mu\nu} $, $h^\text{S}_{\alpha\beta}$ and $a^\text{AdS}_{\mu\nu\rho\sigma}$ are the fluctuation of AdS${}_5$ part of the metric, S${}^5$ part of the metric and  AdS${}_5$ part of the RR 4-form, respectively.
$\Delta=2\ell$ corresponds to the conformal dimension of the operator in the gauge theory.

The classical solution of $s$ with the boundary condition can be written as
% the classical solution of the scalar known as
%\begin{equation}
%s=c\frac{y^\Delta}{K^\Delta}s_0,\:
%K:=|x-x'|^2+y^2.
%\end{equation}
%In more detail,
\begin{equation}
\begin{aligned}
&s(y,x,\theta,\phi,\psi,\cdots)=\int d^4x' c_\Delta \frac{y^\Delta}{K(y,x,x')^\Delta}s_0(x')Y_{\Delta/2}(\psi),\\
&K(y,x,x'):=|x-x'|^2+y^2,\\
&c_{\Delta}=\frac{\Delta+1}{2^{2-\Delta/2}N \sqrt{\Delta}}.\label{classical-solution}
\end{aligned} 
\end{equation}
where $Y_{\Delta/2}$ is the spherical harmonics obtained in appendix \ref{app:harmonics}.  The normalization factor $c_{\Delta}$ is the correct one obtained in \cite{Freedman:1998tz,Lee:1998bxa}.  It is determined so that the coefficient of the two point function is unity.

The first order fluctuation of the action is
\begin{align}
S^{(1)}&=\frac{T_5}{2}\int d^6\zeta \sqrt{\det H}(H^{-1}_{\text{sym}})^{ab}\partial_aX^M \partial_bX^N h_{MN}
    +iT_5\int\mathcal{F}\wedge a_4,
%\nonumber\\
%    &=&T_5\int d^6\xi (\mathcal{L}^{(1)}_\text{DBI}+\mathcal{L}^{(1)}_\text{WZ}),
\label{S1}
\end{align}
where $h_{\mu\nu}$ and $a_4$ are the fluctuation of the metric and the RR 4-form given in eqs.~\eqref{Fluc1}-\eqref{Fluc3}.  $H^{-1}_{\text{sym}}$ denotes the symmetric part of the inverse matrix of $H$.

The one-point function can be calculated by using eq.~\eqref{GKPW1pt}.  The classical action $S_{\text{cl}}$ in eq.~\eqref{GKPW1pt} can be replaced by $S^{(1)}$ in eq.~\eqref{S1}
\begin{align}
 \langle \Ocal(x)\rangle =-\frac{\delta S^{(1)}(s_0)}{\delta s_0(x)}.
\end{align}
The detailed calculation of the fluctuation $S^{(1)}$ is shown in the appendix \ref{app:D5braneAction}.
 The final result of gravity side is given by eq.~\eqref{AppendixGravityResult}
\begin{align}
 -\frac{\delta S_\text{cl}}{\delta s_0(\xi)}
&= C_\ell\frac{\sqrt{\lambda}2^{\ell}\Gamma(2\ell+1/2)}{\pi^{3/2}\sqrt{2\ell}\Gamma(2\ell)} \frac{1}{\xi^{2\ell}}
\int_0^\infty d u 
\frac{u^{2\ell -2}}{\Big[(1-\kappa u)^2+u^2\Big]^{2\ell+1/2}}\label{GravityResult}.
\end{align}
Here $\xi$ is the distance between the interface and the point where the chiral primary operator is inserted.

In eq.~\eqref{GravityResult}, the dependence of $\xi$ is $1/\xi^{2\ell}$ and this is determined by the conformal symmetry. We will compare the coefficient with the gauge theory side in the next section.
\section{Discussion}\label{sec:comparison}
In the previous sections, \ref{sec:gauge} and \ref{sec:gravity}, we calculated the one-point function in the gauge theory side and the gravity side. 
Our goal is to confirm the correspondence between the gauge theory and the gravity theory.
Let us compare these results in this section.
%Taking the limit $\lambda$: fixed, $\lambda/{k^2}\rightarrow 0$, we found the quantities we calculated in the previous sections can be expanded in the power series of $\lambda/{k^2}$.
We consider the limit $k\gg 1$ and $\lambda/k^2 \ll 1$, and compare the leading terms.
\subsection{Gauge theory}
Since we consider the limit $k\gg 1$ the gauge theory result \eqref{gauge-result} becomes
\begin{align}
\left<\mathcal{O}_{2\ell}\right>_\text{classical}
&= C_\ell\frac{(2\pi^2)^\ell}{\sqrt{2\ell}\lambda^\ell}(k^2-1)^{\ell}k\frac{1}{\xi^{2\ell}}\nonumber\\
&\approx C_\ell\frac{(2\pi^2)^\ell}{\sqrt{2\ell}\lambda^\ell}k^{2\ell+1}\frac{1}{\xi^{2\ell}}.\label{approxGauge}
\end{align}
This result is compared with the gravity side.
\subsection{Gravity theory}
We consider the behavior of the gravity side result in the limit $\epsilon:=\frac{1}{\kappa^2+1}\rightarrow 0$, $\kappa = \frac{\pi}{\sqrt{\lambda}}k\gg 1$.
The following expression of the Dirac delta function is convenient\footnote{The case $n=1$ is well known.}.
\begin{equation}
\delta(x)=\lim_{\epsilon\to 0}\frac{1}{\sqrt{\pi}}\frac{\Gamma(n)}{\Gamma(n-1/2)}\frac{\epsilon^{2n-1}}{(x^2+\epsilon^2)^n}.
\end{equation}
Using this formula the integrand of the equation \eqref{GravityResult} can be approximated by the Dirac delta function.
\begin{equation}
\frac{1}{\big((1-\kappa u)^2+u^2\big)^{2\ell+1/2}}\longrightarrow 
\frac{1}{\epsilon^{4\ell}}
\frac{\Gamma(2\ell)\Gamma(\frac{1}{2})}{\Gamma(2\ell+\frac{1}{2})}
\delta(u-\kappa\epsilon).
\end{equation}
After integration we obtain the result
\begin{equation}
-\frac{\delta S^{(1)}}{\delta s_0(\xi)}=
C_\ell \frac{(2\pi^2)^\ell}{\lambda^\ell\sqrt{2\ell}}
k^{2\ell+1}\frac{1}{\xi^{2\ell}}.\label{approxGravity}
\end{equation}
%where we wrote $x_3$ as $\xi$.\\
Comparing \eqref{approxGauge} and \eqref{approxGravity}, we can conclude that these two quantities completely agree in the leading order of $\lambda/k^2$ series.

We can go to next-to-leading order in the gravity side.  Actually the integral in eq.~\eqref{GravityResult} can be rewritten as
\begin{align}
 I:&=\int_0^{\infty}du \frac{u^{2\ell-2}}{[(1-\kappa u)^2+u^2]^{2\ell +1/2}}\nonumber\\
 &=\kappa^{2\ell+1}\left(1+\frac{1}{\kappa^2}\right)^{3/2}
  \int_{-\arctan \kappa}^{\pi/2}d\theta (\cos \theta)^{4\ell-1}\left(1+\frac{1}{\kappa}\tan\theta\right)^{2\ell-2},
\end{align}
by the change of variable as $\tan\theta=(1+\kappa^2)u-\kappa$.  This function can expanded around $\kappa\to \infty$ as
\footnote{This expansion is correct for $\ell\ge 2$}
\begin{align}
 I&=\kappa^{2\ell+1}\frac{\Gamma(2\ell)\Gamma(1/2)}{\Gamma(2\ell+1/2)}
\left(1+\frac{1}{\kappa^2}I_{1}+O(\frac{1}{\kappa^4})\right),\\
I_1&=\frac{3}{2}+\frac{(2\ell-2)(2\ell-3)}{4(2\ell-1)}.
\end{align}
Using this $I_1$ the gravity result up to next-to-leading order is
\begin{equation}
-\frac{\delta S^{(1)}}{\delta s_0(x)}=
C_\ell \frac{(2\pi^2)^\ell}{\lambda^\ell\sqrt{2\ell}}
k^{2\ell+1}\frac{1}{\xi^{2\ell}}
\left(
1+\frac{\lambda}{\pi^2k^2}I_1
+\cdots \right).\label{correction1}
\end{equation}
These corrections are formally a positive power series of $\lambda/k^2$. 
The expansion eq.~\eqref{correction1} indicates the reason why we can compare the gravity side and the gauge theory side.
In the gravity side $\lambda/k^2$ can be small even though $\lambda$ is large because $k^2$ can be larger.    Thus one can suppress the sub-leading terms by sending $\lambda/k^2\to 0$ which has superficially the same effects as $\lambda\to 0$.  A heuristic arguments of $\lambda/k^2$ scaling in the gauge theory side is given in the discussion section of \cite{Nagasaki:2011ue}.

An interesting future work is to compare the prediction of the 1-loop correction in eq.~\eqref{correction1} from the gravity side to the 1-loop calculation in the gauge theory side.
\subsection*{Acknowledgments}
We would like to thank Hiroaki Tanida for discussions and comments.  S.Y. was supported in part by KAKENHI 22740165.

%%%%%%%%%%%%%%%%%%%%%%%%%%%%%%%%%%5
\appendix
%%%%%%%%%%%%%%%%%%%%%%%%%%%%%%%%%%%%%%%%%%%%
\section{Spherical harmonics}\label{app:harmonics}
%\subsection{Hypergeometric function}\label{Hypergeometric}

\subsection{SO(3) $\times$ SO(3) invariant ansatz}\label{InvAnsatz}
The interface in this paper preserves SO(3)$\times $SO(3) symmetry out of SO(6) R-symmetry. Thus only SO(3)$\times$SO(3) invariant operators can have non-vanishing expectation values. 
We would like to introduce SO(3)$\times$ SO(3) invariant spherical harmonics on $S^5$.\\
$S^5$ is described as a hypersurface in 6-dimensional Euclidean space whose coordinates are $(x_4,\dots,x_9)$. $S^5$ is defined by the equation
\begin{equation}
 x_4^2+\dots +x_9^2=1.
\end{equation}
We introduce a parameter $\psi$, $0\le \psi \le \frac{\pi}{2}$ and reexpress this $S^5$ as the following way.
\begin{equation}
x_4^2+x_5^2+x_6^2=\sin^2\psi,\:\:
x_7^2+x_8^2+x_9^2=\cos^2\psi.
\end{equation}
Then the metric is written as
\begin{equation}
ds^2=d\psi^2+\cos^2\psi d\tilde{\Omega}_2^2+\sin^2\psi d{\Omega}_2^2,
\end{equation}
where $d\tilde{\Omega}_2^2$ and $d{\Omega}_2^2$ are line elements of unit $S^2$.

%We seek the form of the Laplacian operator which operates on spherical harmonics.
The SO$(3)\times$SO$(3)$ invariant spherical harmonics only depends on the coordinate $\psi$.
%We assume here spherical harmonics are functions only of a variable $\psi$. 
Let $Y$ be such a function of $\psi$; $Y=Y(\psi)$. The Laplacian operating on this $Y$ is written as
\begin{equation}
\Box Y
=\frac{1}{\sqrt{g}}\partial_i\sqrt{g}g^{ij}\partial_jY
=\frac{1}{\cos^2\psi\sin^2\psi}\frac{d}{d\psi}\cos^2\psi\sin^2\psi\frac{d}{d\psi}Y(\psi).
\end{equation}
After changing the variable $z:=\cos^2\psi$, the Laplacian is rewritten as
\begin{equation}
\Box Y=4z(1-z)\partial_z^2Y+(6-12z)\partial_zY.
\end{equation}
Then the eigenvalue equation, $\Box Y=-EY$, reads
\begin{equation}\label{EigenEqY}
z(1-z)\partial_z^2Y+\bigg(\frac{3}{2}-3z\bigg) \partial_zY+\frac{E}{4}Y=0.
\end{equation}
This is a hypergeometric differential equation.

%Hypergeometric functions play a crucial role in a wide area of mathemaical physics.
%The hypergeometric function is the solution of the hypergeometric function given by
In general a hypergeometric differential equation is given by
\begin{equation}
z(1-z)\partial^2_z F+(c-(a+b+1)z)\partial_zF-abF=0,
\end{equation}
where $a,b,c$ are real parameters.
The solution which is regular at $z=0$ is the hypergeometric function given by an infinite power series
\begin{equation}
F(a,b,c;z)=\sum^{\infty}_{n=0}\frac{(a)_n(b)_n}{(c)_n}\frac{z^n}{n!}.
\end{equation}
Here the Pochhammer symbol $(a)_n=\Gamma(a+n)/\Gamma(a)$ is used.

Since we need the smooth solution on whole $S^5$, 
the solution of eq.~\eqref{EigenEqY} must be regular not only at $z=0$ but $z=1$.
Then the solution must be a hypergeometric function with $a=-\ell,b=\ell+2,c=3/2,\ (\ell=0,1,2,3,\dots)$ and the eigenvalue $E=2l(2l+4)$ is obtained.
% of hypergeometric functions (see section \ref{Hypergeometric}).
Therefore the solution of the equation \eqref{EigenEqY} is expressed in terms of hypergeometric function.
\begin{equation}
Y_\ell(\psi)=C_\ell F(-\ell,2+\ell,3/2;\cos^2\psi),\label{Cl}
\end{equation}
where the normalization factor $C_\ell$ is determined by
\begin{equation}
\int_{S^5}\sqrt{g}|Y_\ell|^2
=\frac{\pi^3}{2^{2\ell-1}(2\ell+1)(2\ell+2)}.\label{Ynormalization}
\end{equation}
The conformal dimension $\Delta$ of the corresponding chiral primary operator is $\Delta =2\ell$.
%And we can also write Spherical harmonics by hypergeometric functions. 
%\begin{equation}
%C^{I_1,I_2,\cdots, I_\Delta}x_{I_1}x_{I_2}\cdots x_{I_\Delta}=Y_\ell (\psi), \: \Delta=2\ell
%\end{equation}

%%%%%%%%%%%%%%%%%%
\section{Detailed calculation}\label{app:D5braneAction}
%%%%%%%%%%%%%%%
\subsection{Fluctuations $h$ and $a$}
In this appendix we show the detailed calculations of fluctuations $h$ and $a$ defined by the scalar field $s(x)$ as \eqref{Fluc1},\eqref{Fluc2} and \eqref{Fluc3}.
Actually it is enough to calculate them when $s_0$ is a delta function as
\begin{align}
 s_0(x)=\delta^{4}(x-x').\label{sdelta}
\end{align}
In this case the classical solution \eqref{classical-solution} becomes
\begin{align}
 s(y,x,\theta,\phi,\psi)=c_{\Delta}\frac{y^{\Delta}}{K(y,x,x')^{\Delta}}Y_{\Delta/2}(\psi).
\end{align}

We use the convention for the covariant derivative and totally anti-symmetric tensor
\begin{eqnarray}
&&\nabla_i T_{j_1\cdots j_n}:=\partial_i T_{j_1\cdots j_n} -\sum^n_{l=1}\Gamma_{ij_l}^kT_{j_1\cdots j_{l-1}k j_{l+1}\cdots j_n}, \\
&&\epsilon_{y0123}=1,
\end{eqnarray}
where Christoffel symbols are $\Gamma^i_{jk}:=\frac{1}{2}g^{il}(\partial_j g_{lk}+\partial_k g_{lj}-\partial_l g_{jk})$.

The first derivatives and the second derivatives of $s$ are
\begin{align}
\frac{\partial_ys}{s}&=\Delta(\frac{1}{y}-\frac{2y}{K}),\\
\frac{\partial_is}{s}&=-\Delta\frac{2(x-x')_i}{K},
\end{align}
\begin{align}
&\frac{\nabla_y\nabla_ys}{s}=\frac{\Delta^2}{y^2}+4\Delta(\Delta+1)\left(
	-\frac{1}{K}+\frac{y^2}{K^2}\right),\\
&\frac{\nabla_y\nabla_is}{s}=\Delta(\Delta+1)\left(
	+4y\frac{(x-x')_i}{K^2}-2\frac{(x-x')_i}{yK}\right),\\
&\frac{\nabla_i\nabla_js}{s}=-\Delta\frac{\delta_{ij}}{y^2}
	+4\Delta(\Delta+1)\frac{(x-x')_i(x-x')_j}{K^2}.
\end{align}
Using these results and the definition of $h$
in AdS the expression of fluctuations are
\begin{align}
&\frac{h^\text{AdS}_{yy}}{\Delta s}
	=\frac{2}{y^2}-\frac{16}{K}+\frac{16}{K^2},\\
&\frac{h^\text{AdS}_{yi}}{\Delta s}
	=16y\frac{(x-x')_i}{K}-8\frac{(x-x')_i}{yK},\\
&\frac{h^\text{AdS}_{ij}}{\Delta s}
	=-2\frac{\delta_{ij}}{y^2}+\frac{16(x-x')_i(x-x')_j}{K^2},
\end{align}
and in 2-sphere 
\begin{align}
\frac{h^\text{S}_{\theta\theta}}{\Delta s}=2,\:\:
\frac{h^\text{S}_{\phi\phi}}{\Delta s}=2\sin^2\theta.
\end{align}

%%%%%%%%%%%%%%%%
\subsection{D5-brane action}
When we give fluctuation to the metric and the RR 4-form,
the D5-brane action is deformed as follows in the first order.
We use the notation $v_i=x_i-x'_i$ and $p,q$ run $0,1,2$.
The first order fluctuation is calculated as follows.
\begin{eqnarray}
S^{(1)}&=&\frac{T_5}{2}\int d^6\zeta \sqrt{\det H}(H^{-1}_{\text{sym}})^{ab}\partial_aX^M \partial_bX^N h_{MN}
	+iT_5\int\mathcal{F}\wedge a_4\nonumber\\
	&=&T_5\int d^6\zeta (\mathcal{L}^{(1)}_\text{DBI}+\mathcal{L}^{(1)}_\text{WZ}).
\label{DBI+WZ(1)}
\end{eqnarray}
In this equation we need the explicit form of the symmetric part of $H^{-1}$.
\begin{align}
 H^{-1}_{\text{sym}}=
\begin{pmatrix}
 (1+\kappa^2)^{-1}y^{2}&&&&&\\
 &y^{2}&&&&\\
 &&y^{2}&&&\\
 &&&y^{2}&&\\
 &&&&(1+\kappa^2)^{-1} &\\
 &&&&& [\sin^2\theta(1+\kappa^2)]^{-1}
\end{pmatrix}.
\end{align}
Eq.~\eqref{DBI+WZ(1)} is calculated as follows.
\begin{align}
\mathcal{L}^\text{(1)}_\text{DBI}
:=&\frac{1}{2}\sqrt{\det H}(H^{-1}_{\text{sym}})^{ab}\partial_aX^M\partial_bX^Nh_{MN}\nonumber\\
=& \frac{(1+\kappa^2)\sin^2\theta}{2y^4}
	\{H^{yy}\partial_y X^M \partial_y X^N h^\text{AdS}_{MN}
		+H^{ij}\partial_i X^M \partial_j X^N h^\text{AdS}_{MN}\nonumber\\
&\qquad		+H^{\theta\theta}\partial_\theta X^M \partial_\theta X^N h^\text{S}_{MN}
		+H^{\phi\phi}\partial_\phi X^M \partial_\phi X^N h^\text{S}_{MN}
		\}\nonumber\\
=& \frac{\Delta s\sin\theta}{y^4K^2}\{-8y^2v_3^2+\kappa(16y^3v_3-8yv_3K)+\kappa^2(8y^2(v_pv_p+v_3^2)-4K^2)\}.
\end{align}

\begin{align}
\mathcal{L}^\text{(1)}_\text{WZ}
:=&i\mathcal{F}_{\theta\phi}\frac{1}{4!}\epsilon^{abcd}(Pa)_{abcd}\nonumber\\
=& i2\kappa \sin\theta(a_{y012}+\kappa a_{3012})\nonumber\\
=& i2\kappa\sin\theta\left\{
	\Delta 4s\frac{1}{y^3}\frac{2v_3}{K}
	+\kappa\Delta 4s\frac{1}{y^3}	
		\Big(\frac{1}{y}-\frac{2y}{K}\Big)
	\right\}\nonumber\\
=& \frac{i\sin\theta\Delta s}{y^4K^2}
	\left\{\kappa(16v_3yK)+\kappa^2(8\kappa^2-16y^2K)\right\}.
\end{align}
$S^\text{(1)}$ is the sum of these two terms
\begin{eqnarray}
S^\text{(1)}
&=& T_5\int d^6 \zeta	\Big(\mathcal{L}^\text{(1)}_\text{DBI}+\mathcal{L}^\text{(1)}_\text{WZ}\Big)\nonumber\\
&=& -8T_5\int d^6 \zeta
	\frac{\sin\theta\cdot \Delta s}{y^2K^2}(v_3-\kappa y)^2\nonumber\\
&=& -8T_5\int d^6 \zeta	
	\frac{\sin\theta\cdot \Delta s}{y^2K^2}x_3'^2.
\end{eqnarray}
%\begin{eqnarray}
%&&\mathcal{L}^{(1)}_\text{DBI}
%=\frac{\sin\theta\cdot\Delta s}{2y^2}
%	\left\{
%	-16\frac{v_3^2}{K^2}+\kappa \left(-16\frac{v_3}{y\cdot K}+32y\frac{v_3}{K}\right)+\kappa^2\left(-\frac{8}{y^2}+16\frac{v_pv_q+v_3^2}{K^2}\right)
%	\right\}\nonumber\\ \\
%&&\mathcal{L}^{(1)}_\text{WZ}
%=\kappa\sin\theta\frac{4\Delta s}{y^3}
%	\left(\frac{2v_3}{K}+\kappa\left(\frac{1}{y}-\frac{2y}{K}\right)
%	\right)
%\end{eqnarray}
%We show here the result.
%\begin{equation}
%S^{(1)} = -8T_5\int d^6\xi \frac{\Delta s\sin\theta}{y^2K^2}(\kappa y-v_3)^2.
%\end{equation}
This formula with the classical solution \eqref{sdelta} $s_{0}(x)=\delta^4(x-x')$ is the functional derivative $\delta S^{(1)}/\delta s_0(x')$.  This functional derivative evaluated at $x_3'=\xi$ is the quantity we want.  Notice that the D5-brane sits at $\psi=\pi/2$, thus the spherical harmonics should be evaluated at this surface.  This value is given by (see eq.~\eqref{Cl})
\begin{align}
 Y_{\ell}(\psi=\pi/2)=C_{\ell}.
\end{align}
Putting all these things together, we obtain
\begin{align}
-\frac{\delta S^{(1)}}{\delta s_0(\xi)}
=& 32T_5\pi\Delta c_\Delta C_\ell \int^\infty_0dy\int dx^0dx^1dx^2
	\frac{y^{\Delta-2}\xi^2}{((\kappa y-\xi)^2+x^p x^p+y^2)^{\Delta+2}}\nonumber\\
=& 32T_5\pi^{5/2}\Delta c_\Delta C_\ell \frac{\Gamma(\Delta+1/2)}{\Gamma(\Delta+2)}
	\xi^2\int^\infty_0dy \frac{y^{\Delta-2}}{((\kappa y-\xi)^2+y^2)^{\Delta+1/2}}.
\label{Integrated}
\end{align}
In the above calculation we used the formula.
\begin{equation}
\int d^Dx\frac{1}{(x^2+A)^\alpha}=\frac{\Gamma(-D/2+\alpha)}{\Gamma(\alpha)}\frac{\pi^{D/2}}{A^{-D/2+\alpha}}.
\end{equation}
%Let $x_3=\xi$ for comparison to gauge theory.
In our unit \eqref{unit} the D5-brane tension is written as $T_5=\frac{2N\sqrt{\lambda}}{(2\pi)^4}$.
Finally by substituting $T_5$, $c_{\Delta}$ and $\Delta=2\ell$ to eq.~\eqref{Integrated}, and the change of valuable as $y=\xi u$, we obtain
\begin{align}
 -\frac{\delta S_\text{cl}}{\delta s_0(\xi)}
&= C_\ell\frac{\sqrt{\lambda}2^{\ell}\Gamma(2\ell+1/2)}{\pi^{3/2}\sqrt{2\ell}\Gamma(2\ell)} \frac{1}{\xi^{2\ell}}
\int_0^\infty du 
\frac{u^{2\ell -2}}{\Big[(1-\kappa u)^2+u^2\Big]^{2\ell+1/2}}\label{AppendixGravityResult}.
\end{align}
%In the last line we used the relation $\kappa=\frac{\pi}{\sqrt{\lambda}}k$ and changed the variables $u=\frac{y}{x_3}$.
\providecommand{\href}[2]{#2}\begingroup\raggedright\endgroup

\end{document}